\documentclass[12pt,preprint]{aastex}

\newcommand\be{\begin{equation}}
\newcommand\ee{\end{equation}}
\shortauthors{H. M. Antia}
\shorttitle{Does the Sun shrink?}
\received{2002 March 21}
\begin{document}

\title{Does the Sun shrink with increasing magnetic activity?}
\author{H. M. Antia}
\affil{Tata Institute of Fundamental Research,
Homi Bhabha Road, Mumbai 400005, India}
\email{antia@tifr.res.in}

\begin{abstract}
It has been demonstrated that frequencies of f-modes can be used to
estimate the solar radius to a good accuracy. These frequencies have
been used to study temporal variations in the solar radius with
conflicting results.
The variation in f-mode frequencies is more
complicated than what is assumed in these studies. If a careful
analysis is performed then it turns out that there is no evidence
for any variation in the solar radius.

\end{abstract}

%\keywords{Sun: oscillations --- Sun: activity}

\section{INTRODUCTION}

Temporal variations in the solar radius has been a controversial topic, as
direct measurements of the solar radius have given conflicting results
(Delache, Laclare \& Sadsaoud~1985; Wittmann, Alge \&
Bianda~1993;
Fiala, Dunham \& Sofia~1994; Laclare et al.~1996; No\"el
1997; Emilio et al.~2000).
The reported temporal variations in the solar radius ranges from 0 to 700
km.
It is important to estimate the radius variations with solar cycle
as these can provide useful
constraint for models to explain the luminosity variation with solar
cycle (Gough 2001). In particular, the ratio of the radius variation to the luminosity
variation, $W=(\Delta R/R)/(\Delta L/L)$ depends on the theoretical
model of luminosity variations. The luminosity variation is known
to be about 0.001 (e.g., Mecherikunnel 1994) between the maximum
and minimum of solar activity.
Thus, it is important to obtain a
reliable estimate of radius variation over the solar cycle
so that we can distinguish between these models.

Recently, Schou et al.~(1997) and Antia (1998) have demonstrated that
the frequencies of f-modes can be used to estimate the solar radius.
Since these frequencies have been measured with a relative accuracy
of $10^{-5}$ we may expect to determine the solar radius to similar
accuracy. However, there are systematic errors of order of 100 km
in calibration of photospheric radius from the measured
frequencies (Tripathy \& Antia 1999).  If these
systematic errors are independent of time, then it would be possible
to determine the temporal variation in the solar radius using f-mode
frequencies. The attempts so far (Dziembowski et al.~1998, 2000, 2001;
Antia et al.~2000, 2001) give conflicting
results. Using the first few data sets from the Michelson Doppler Imager
(MDI), Dziembowski et al.~(1998) found that the solar radius is
increasing with solar activity. They found an increase by about
4 km in 6 months just after the solar minimum. If this variation
was indeed correlated to solar activity we would expect a much
larger variation in radius during the solar cycle.
Subsequently, using more data
Dziembowski et al.~(2000) found no systematic variation in the solar
radius. This work used all data sets from MDI that were obtained
before the contact with SOHO satellite was lost. Using a few data sets from
the Global Oscillation network Group
(GONG), Antia et al.~(2000) found the solar radius to be decreasing with
activity, but subsequently using more extensive data sets from
GONG and MDI Antia et
al.~(2001) found no evidence for any variation in the solar radius.
However, using essentially same data sets from MDI, Dziembowski et al.~(2001)
(hereinafter DGS) have found a decrease in the solar radius.
Unfortunately, the claimed variation, if any, in all these works
is of the order of a
few km and even a small change in systematic errors can give
rise to spurious variations of this order. Clearly, a more careful
analysis of f-mode frequencies is required before drawing any conclusions
about variation of the solar radius.

Antia et al.~(2001) have shown that the variation in f-mode
frequencies is more complex than what is assumed in other studies.
These variations can be decomposed into at least
two components. One of these components is oscillatory with a period of
1 yr, while the second
component is correlated with solar activity. The amplitudes
of both these components increase with frequency and hence are not
likely to arise from radius variations. Variation in the solar radius will
cause frequency shifts that are proportional to frequency, but the
observed variations have much steeper dependence on frequency.
The oscillatory
component is most likely to be an artifact introduced by orbital
period of the Earth. Antia et al.~(2001) have also shown that, most
of the discrepancy between different results about radius variation
using f-mode frequencies
can be explained if these two components are invoked in the temporal
variations. In particular, Antia et al.~(2000) failed to detect
the oscillatory component as they used only 5 data sets covering
a period of 3 years.
Further, after accounting for these two components in
temporal variations there is no evidence for any variation
in the solar radius.
DGS have claimed that the
solar radius decreases at a rate of 1.5 km yr$^{-1}$ during 1996--2000.
However, they have not removed the oscillatory component in f-mode
frequency variation and hence their claim needs to be examined
carefully.

\section{RADIUS VARIATION FROM F-MODE FREQUENCIES}

It can be easily shown that if the solar radius varies by even 1 km
over the solar cycle, the rate of resulting release or absorption
of gravitational energy
would be larger than the solar luminosity. Hence, we can rule out
such radius variations. Thus, any possible variation in the solar radius
must be confined to the outermost layers of the Sun.
DGS have argued that since observed f-modes
are trapped in a layer beneath the
visible surface, they would measure the radius variation at this depth.
In particular, the fractional variation in radius, could be a function
of radial distance. Such a variation is, of course, realistic, but the problem
is, it may not be possible to analyze such variations easily.
For example, DGS have split the f-mode frequency
variations into two parts, one arising from radius variation and the
other from some variations in the outermost layers, which scales
inversely as the mode inertia.
Thus, they express the change in f-mode frequencies as
\begin{equation}
\Delta \nu_\ell=-{3\over 2}{\Delta R\over R}\nu_\ell
+{\Delta\gamma\over I_\ell}\;,
\end{equation}
where $\nu_\ell$ is the frequency of the f-mode of degree $\ell$,
$\Delta R$ is the change in radius while
$\Delta\gamma$ measures the contribution from surface term and
$I_\ell$ is the mode inertia.
While fitting the expression to the observed data DGS assume
$\Delta R/R$ to be constant, which implies that the radius
variations are homologous, at least, in the region where the observed modes
are trapped. Thus as far as f-modes are concerned they have assumed
that $\Delta R/R$ is constant, presumably because otherwise it is
difficult to proceed with the analysis.
Subsequently, they claim that these radius variations  
arise from magnetic field variation in a layer below the
outermost surface layers. This is certainly conceivable,
but if that is the case then there should be an additional term in
Eq.~(1) that arises from the direct effects of the magnetic field.
The effect of the magnetic field on f-mode frequencies cannot be 
assumed to be solely due to 
those arising from radius variation. Frequency shifts
due to magnetic fields (e.g., Campbell \& Roberts 1989)
are not in general, proportional to frequency
as is implied by Eq.~(1) (the surface term cannot arise from such fields in the interior).
Thus, a more
complex model will be required to fit the frequency differences
arising from magnetic field. The same applies to frequency
variations due to density perturbations (Chitre et al.~1998).
Basically, if f-mode frequency
variations are due to magnetic field or density perturbation, then
we need to calculate these shifts explicitly, rather than modeling
them via radius variation, which cannot account for the entire effect.

To estimate the depth at which f-modes are trapped, we can consider
the kinetic energy density from the eigenfunctions of f-modes in
the relevant range of $\ell=140$--300 which are mainly used in this study.
If we assume that the trapping region of each mode covers the layers
where the kinetic energy density is greater than $1/e$ of its peak value, then
the upper limit of $\ell=300$ f-mode is at a depth of about 1 Mm,
while the lower limit of $\ell=140$
f-mode extends to a depth of about 12 Mm. Hence, a depth range of
1--12 Mm is expected to be covered by this study.
The f-modes
are not trapped between a pair of rigid boundaries and the extent of region
covered by them will depend on the definition of boundary as well as on the
precise mechanism responsible for frequency variations.

\section{RESULTS USING MDI DATA}

The inconsistencies pointed out in the previous section arise
in attempting to find a physical model which gives rise to the
radius variation inferred from f-mode frequencies. This
model is relevant only if the data actually show any evidence for change in
radius from f-mode frequencies. Thus, in this section,
I ignore the cause of radius variation and just address
the question whether the changes in f-mode frequencies imply any change in
the solar radius (as defined by DGS).
The f-mode frequency variation is expressed using two terms,
one arising from variation in the solar radius
and another from  unspecified variations in the outermost surface layers.
There are good reasons to look for such a term since
it is known that p-mode frequency variations largely arise from
variations in outer layers (e.g., Basu \& Antia 2000).
Thus, following DGS, I assume the frequency variation to be given by 
Eq.~(1), where $\Delta R/R$ and $\Delta \gamma$ are constants.
In order to test whether the observed data fit this form, I have used the same
f-mode frequency data (Schou 1999) that DGS have used except for
some additional data sets that are now available.
These data consist of 30 sets each covering a period of 72 days
starting from May 1, 1996 and ending on August 21, 2002.
Note that there is a gap in data sets between 1998.5 and 1999.2,
when the contact with SOHO satellite was lost. The only data set
during this period has significantly worse fit and may be ignored.
For each data set, I take the difference
in frequency with respect to a standard solar model (in the
sense observed $-$ model) with radius
$R_\odot=695.78$ Mm. The model radius is chosen to ensure that
the frequency differences are small. 
These frequency differences are then fitted to
Eq.~(1). All modes with $\ell>140$ are used in these fits.
Even if I ignore the data set taken immediately
after recovery of SOHO satellite ($\chi^2=3.54$), the $\chi^2$ per degree of freedom
in these fits varies from 1.2 to 2.5. One such fit for data obtained
around 1997.0 is shown in Fig.~1.
It is clear that the fit is not good and that the variation in
frequency differences is more complicated than what is modeled
by Eq.~(1). This should be expected from the results of
Antia et al.~(2001), since depending on the phase of the oscillatory
component in the frequency
variations, it can have a different sign as compared to the other component of the variation.
Furthermore, the oscillatory component has roughly the same frequency
dependence as the second term in Eq.~(1), while the non-oscillatory
component has a significantly less steep frequency dependence and hence,
may not be adequately represented by Eq.~(1).
Thus, at times when both components are in phase the net frequency
variation may be approximated by Eq.~(1), but six months later when
the oscillatory component changes sign the two will be opposite, as is
the case around 1997.0, data for which are shown in Fig.~1.
Since the oscillatory component
has a steeper frequency dependence as compared to the non-oscillatory
component, at high frequencies the trend appears
to reverse. Such a behavior cannot be modeled by Eq.~(1).

Fig.~2 shows the results from fits to all data sets from MDI.
The upper panel shows the inferred radius variation, which is
similar to Fig.~2 of DGS. The middle panel shows the fitted
variation in surface term $\Delta\gamma$. This figure also looks
similar to Fig.~2 of DGS, though the y-axis is different. The
cause of this difference is not clear as the surface term cannot
be positive as shown in Fig.~2 of DGS. The lowest panel shows
the $\chi^2$ per degree of freedom for each of the fits.
The oscillatory trend is quite clear in all these panels.
Further, comparing different panels it is clear that the best fits
are obtained when the oscillatory and non-oscillatory components
are of the same sign and the magnitude of the oscillatory
component is close to maximum, which is the case when $\Delta\gamma$ is the
lowest. The non-oscillatory
component appears to be reducing with increasing solar activity
and as a result, fits improve during high activity period.
For example, the best fits during pre-recovery period have
$\chi^2\approx 1.5$ per degree of freedom, while during 2000, 2001
this reduces to 1.2. 
In their Fig.~1, DGS have shown fits to some data sets at 
interval of 1 yr. It can
be seen from Fig.~2 in this paper that these correspond to times
when $\chi^2$ is close to a local minima.
The fits shown by DGS correspond to filled squares in Fig.~2.
Some of the intermediate data sets give bad fits as can be seen
from Fig.~1.
About 20\% of the fits have $\chi^2\ga 2$ and most of these show
clear deviation from the assumed form.

These oscillations in $\Delta R$ arise because the expression is inadequate
to fit the data and do not represent real variations in $\Delta R$.
Basically there is additional contribution to $\delta\nu_\ell$ which
can not be represented by either of the terms in Eq.~(1) and this
gets projected on to the two terms giving spurious results.
In particular, the oscillatory trend is also projected on to both
terms in Eq.~(1) and further, the division between the two terms
is also a function of time. As a result, the oscillations get
modulated and it is not straightforward to isolate the oscillatory
part in the fitted results for $\Delta R$.
If all these oscillations are ignored and a
linear function in time is fitted to the inferred radius variation as DGS have done,
then the
$\chi^2$ per degree of freedom for this fit is 8.7.
This fit is shown by the continuous line in the
upper panel of Fig.~2 and corresponds to a radius variation of 
$-1.2$ km yr$^{-1}$, slightly less than that inferred by DGS. This difference
is because of the additional data that have become available.
A slightly smaller $\chi^2$ of 5.5 per degree of freedom
is obtained if instead, these points are fitted by
a step function with a discontinuity around 1999.2. This fit is
shown by heavy line in upper panel of Fig.~2. Looking at the
top panel of Fig.~2, it appears that inferred radius has suddenly changed
around 1999 and the step function fit appears to support this
hypothesis.
The large $\chi^2$ is to be expected as Eq.~(1) that is used to calculate
$\Delta R$, does not really fit the observed data at all times.
%Hence there are
%other terms in frequency variation which need to be included in a
%realistic analysis.

Ideally, one should remove the
oscillatory component in frequency variation before considering longer
period variations, but for simplicity, I consider fits at interval of
1 year which will be at the same phase of oscillatory component
and further, select the phase such that the fits are the best in some sense.
These points are marked by filled squares in Fig.~2.
Table~1 gives the results obtained for these sets, which includes
the $\chi^2$ per degree of freedom as well
as the average 10.7 cm radio flux during the time interval covered by
the data set, which is a measure of solar activity.
Looking at this table
it is clear that the radius is not changing continuously. In fact,
most of the radius variation has occurred between 1998.4 and 1999.4.
Possible radius variation during 1996.4--1998.4 and 1999.4--2002.4 is less than
1 km.
The solar activity did increase significantly during the period 1998.4--1999.4,
but there has been comparable change in activity during other
periods too. Hence that cannot explain the variation seen in Table~1.
This happens to be the period during which contact with SOHO satellite was lost
and it is most likely that this variation reflects systematic
errors arising from changes in the MDI instrument that may have occurred
during recovery of the satellite. Even if
we assume that this variation is real, the rate of shrinking is
not 1.5 km yr$^{-1}$ as claimed by DGS, but something like
3 km yr$^{-1}$ during 1998.4--1999.4 and essentially no variation at
other times. Thus, any model to explain this frequency change
by a radius variation must
explain why there is little radius variation during most of the time and
why all variation is confined to less than 1 yr at some intermediate
phase of solar cycle.

In order to study the robustness of the inferred radius variation,
I attempt the fits by restricting the mode set or the data sets and
the results are summarized in Table 2.
If high degree modes are neglected, then the fits to data using Eq.~(1)
improve to some extent, which is mainly because the total variation
in frequencies reduces with degree. Nevertheless,
the fit to linear
variation in $\Delta R$ is still bad and its slope keeps reducing
as the upper limit on $\ell$ is reduced. Thus if only modes with
$140<\ell<250$ are used the radius variation comes out to be
$-0.57\pm0.08$ km yr$^{-1}$ (with a $\chi^2=2.6$),
while if the upper limit on $\ell$
is reduced to 200, it becomes $-0.47\pm0.17$ km yr$^{-1}$
($\chi^2=2$). In these cases if a step function is fitted the
$\chi^2$ comes out to be 2.1. Figure 3 shows the fits in these
cases. It can be seen that the magnitude of possible discontinuity
around 1999 reduces as the upper limit on $\ell$ is reduced and
is hardly visible when the upper limit is reduced to $\ell=200$.
In this case the errors in inferred radius are rather large and
the one year oscillations are
essentially wiped out by statistical fluctuations.
The reduction in $\chi^2$ is mainly due to increase in estimated
errors in $\Delta R$.
Antia et al.~(2001) have shown that
the amplitude of oscillatory term reduces with decreasing $\ell$
and that also contributes to improvement in fits.
If the $\ell$ range is reduced still further, the errors in fitted
quantities are too large to make any meaningful deductions and the results
are not shown in Table~2. This mainly arises because the frequency
variation are very small and it is not easy to distinguish
between contributions of the two terms in Eq.~(1) over a small range
of $\ell$.

If only the first term in Eq.~(1) is used
for fitting, then it is equivalent to taking an average of
relative frequency variations over all modes. In that case,
any variation in frequency will imply a variation in radius.
This may not be realistic and the resulting fits (to Eq.~(1)) are always
bad. Since both components in the frequency variations increase
steeply with $\ell$, as the upper limit on $\ell$ is reduced
the inferred radius variation should reduce and the limiting
value at the lowest frequency range would give an upper limit
to any possible radius variation.
In this case,  since only one parameter is fitted, it is possible
to get some fits with only a few low degree modes and hence it is
possible to reduce the upper limit on $\ell$. 
The estimated rate of reduction in radius decreases from 2.2 km yr$^{-1}$
when all modes are used to 0.74 km yr$^{-1}$ when the upper limit
on $\ell$ is reduced to 160. If the upper limit on $\ell$ is
reduced still further, there are very few modes in some data sets
and it is not possible to obtain any meaningful fits.
However, recently the MDI data sets have been updated and the
new data has more f-modes. With these revised data sets it is possible to
reduce the upper limit to $\ell=120$ and the
inferred radius variation comes out to be $+0.08\pm0.11$ km yr$^{-1}$
(Fig.~4).
As demonstrated by Antia et al.~(2001), at these low frequencies
the oscillatory component in frequency
variation is also not observed.
This fit appears to be consistent with results of Basu \& Antia (2002)
who found that systematic error in MDI data is restricted to
modes with $\ell>120$. When these modes are eliminated no
radius variation is found.
In order to enable a direct comparison with DGS, these revised data
sets are not generally used in this work, but Table~2 lists the results
obtained using these sets also over the full time interval.
It is clear that the results are not significantly different from
earlier results.
If the fits are restricted to include
only the post gap data sets (i.e., after 1999) and still use
only the first term in Eq.~(1), then the resulting
rate of radius decrease is 1.1 km yr$^{-1}$ with all modes and
comes down to $0.03\pm0.22$ km yr$^{-1}$ when modes with $\ell<160$
are considered. This is to be expected as the frequency variation
increases rapidly with degree. Thus the inferred radius variation
is maximum when high degree modes are used and is negligible when
only relatively low degree modes are used. If there is any
component in frequency differences which corresponds to radius
variation, the limiting radius variation will tend to this value
when the modes in low $\ell$ range are used. Since this limiting
value happens to be consistent with zero, we can conclude that there
is no radius variation during 1999--2002 (or during the entire period
when the revised MDI data are used).
On the other hand, if the second term in Eq.~(1) is
also included and only data sets after 1999.0 are used then
the resulting fit does not show any significant variation in radius
irrespective of the upper limit on $\ell$. Similar results are
obtained when only data sets before the gap (i.e., before 1998.6)
are used.  Thus it is clear that
most of the inferred variation in solar radius has taken place
during the gap in MDI data. 

From the top panel in Fig. 2 it can be seen that in the pre-gap
data (before 1999.0) the filled squares are close to the minimum
in $\Delta R$, while after the gap the filled squares are
close to the maximum in $\Delta R$. The reason for this flip is
not clear. It could be due to instrumental variations during
recovery or alternately it may be because before the gap the
non-oscillatory component had larger amplitude as compared to
the oscillatory component, while after the gap the amplitude of
non-oscillatory component is less. There is also some variation in
the number of modes and the set of modes between different data sets.
If the expression fits the data well this variation will not matter,
but unfortunately that is not true.
It is difficult to assign
much significance to these results as during the time of these
data sets oscillatory component has maximum magnitude and this
can give rise to spurious results in the fits.
Comparing Figs.~2 and 4 it can be seen that
the inferred radius using $\ell<120$ modes is in between the
pre-gap and post-gap values using all modes. Thus it is clear
that averaging over the oscillations in Fig.~2 does not give the
correct estimate of radius.
Ideally, one should
subtract out the oscillatory component in the frequency variation
itself before analyzing the data, as has been done by
Antia et al.~(2001).

\section{DISCUSSION AND CONCLUSIONS}

There have been a few claims in the recent times about variation in
the solar radius from f-mode frequencies (Dziembowski et al.~1998, 2001;
Antia et al.~2000). Unfortunately, the 
variation in f-mode frequencies is more complex than what is
modeled in these studies.
From the results shown in Table~2, it is clear that there is no
evidence to support a
decrease in radius since the result depends on the set of modes included in
the study and the linear fit is generally bad. From Table~2 it
can be seen that the estimated rate of radius change varies between
$-2.2$ and $+0.1$ km yr$^{-1}$ depending on the range of $\ell$
and the number of terms in Eq.~(1) used for fitting.
This large variation is simply because Eq.~(1) does not fit the
observed data properly. More terms will be required to obtain a
proper estimate of radius variation.
In particular, when
only $\ell<120$ f-modes are used the inferred radius variation
always comes out to be consistent with zero.
Similarly, when only data sets before (or after) the gap are used
the inferred radius variation comes out to be negligible.
Basically, the observed data sets do not
appear to have any component of relative frequency variation
which is independent of degree as would be required for radius
variation (see Eq.~(1)).

A large part  of the inferred  variations in solar radius is most
probably due to instrumental effects. For example,
oscillatory trend with a period of 1 year is probably due to orbital
motion of the Earth and SOHO satellite around the Sun.
Similarly, the sharp variation seen between 1998.4 and 
1999.4, (depending on degree $\ell$) is most likely a result
of changes in instrumental characteristics during the recovery of
the SOHO satellite. These two effects can account for all claims
of radius variation made earlier.
All these
instrumental errors need to be eliminated before any claim can
be made about the cause of frequency variation.
From the results presented above
it is clear that after eliminating these instrumental
effects there is no significant variation in
the solar radius as determined by f-mode frequencies.
Similar conclusion was obtained by Antia et al.~(2001) using a more
detailed analysis of both GONG and MDI data.

The systematic error between MDI data sets before and
after recovery also manifests in other studies (Antia 2002;
Basu \& Antia 2002; Antia, Chitre \& Thompson 2003).
In particular, it is found that this systematic error is mostly
confined to modes with $\ell>120$, which is consistent with the
results in this work. When these modes are neglected no radius
variation is found, while if these are included then we find
varying amount of radius variation around 1999.
If the radius variation is real, it cannot depend on $\ell$.
It is quite likely that systematic errors are present in all
MDI data sets, but their magnitude has changed during recovery.

If we assume that the inferred radius variation between
1998.4 and 1999.4 is of instrumental origin, then we can put
some limits on radius variation.
From Table 1 it can be seen that the maximum variation between the 3 points
before the data gap is 1.3 km, while that in the 4 points after
the gap is 1.1 km. Considering an error of about 0.6 km in each
data point, this variation is consistent with no radius variation.
This would suggest an upper limit comparable to error bars in each point
on radius variation during half of the solar cycle. Similar conclusion
can be obtained from Fig.~4, which shows the results using only
$\ell<120$ modes which are not expected to be affected by the
systematic error in MDI data. From these results we can put a
conservative upper limit of 2 km on radius variations during the
last 6 years. This would yield
 $\Delta R/R < 3\times10^{-6}$ and
$W<0.003$ as the ratio of radius to luminosity variation.
Such a small value should favor models involving changes in the
outer layers to explain the observed luminosity variations
(Gough 1981; D\"appen 1983; Balmforth, Gough \& Merryfield 1996;
Gough 2001). Of course, the value of $W$ in these models is
determined by radius variation at the photosphere, while f-modes
are sensitive to variations at depths of about 1--12 Mm.
But if the photospheric radius variations are much larger than
those inferred by the f-modes, then the cause is most likely to be
near the surface. Emilio et al.~(2000) find a much larger increase
in photospheric radius by about $15\pm2$ km during the solar cycle.
These direct measurements from MDI are also affected by
a number of systematic errors and as they have pointed out this value
should be regarded as an upper limit to radius variation. Thus
our results from f-mode frequencies which effectively measure the
solar radius in the subsurface layers, are probably not inconsistent with
these measurements.

\section*{ACKNOWLEDGMENTS}

This work  utilizes data obtained by the Solar Oscillations
Investigation / Michelson Doppler Imager  on the Solar
and Heliospheric Observatory (SOHO). SOHO is a project of
international cooperation between ESA and NASA.

\clearpage

\begin{figure}
\plotone{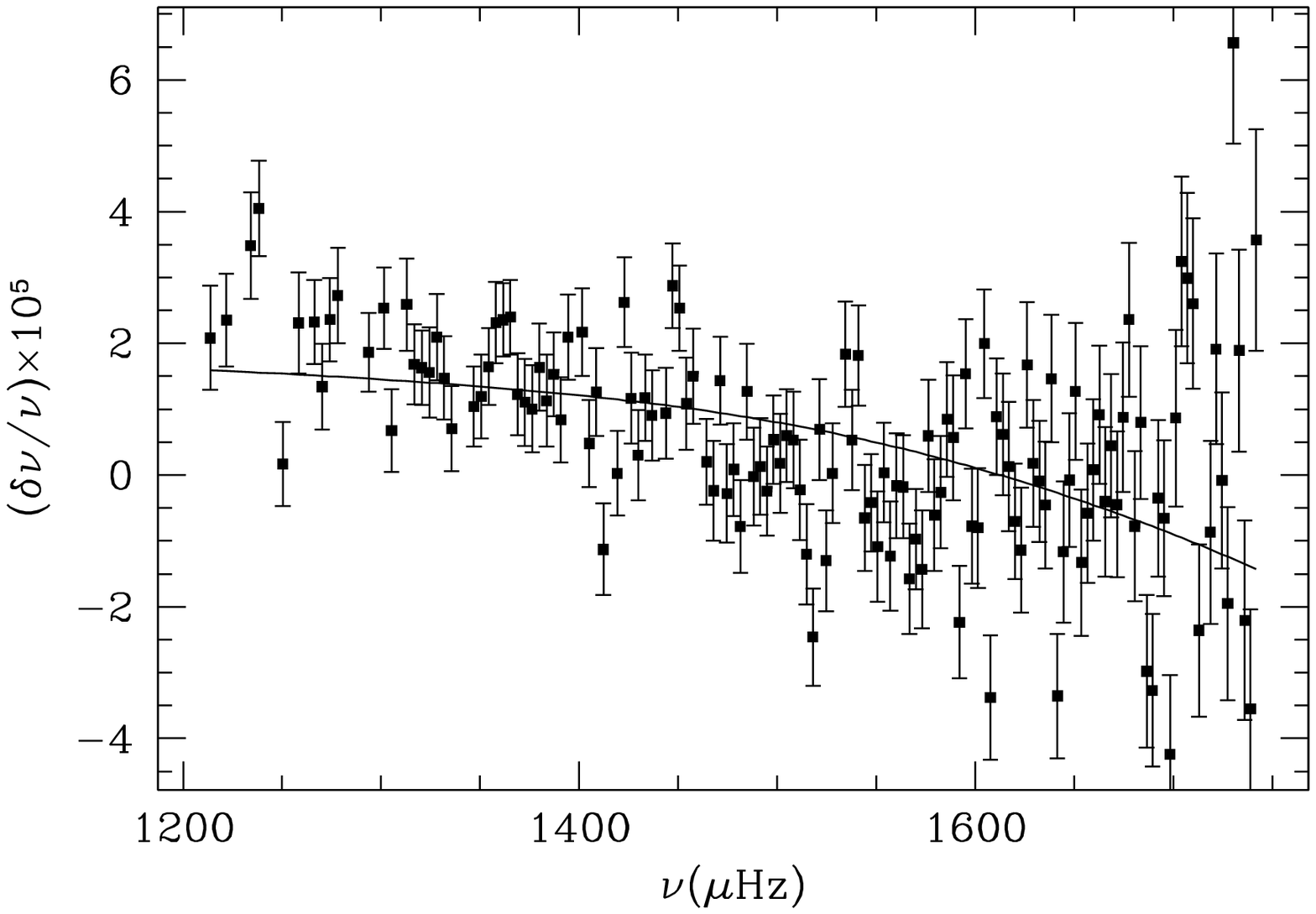}
\caption{Fit to MDI data taken around 1997.0 using Eq.~(1).
The points with error bars show the relative frequency difference
between observed and model frequencies. The continuous line shows the
best fit using Eq.~(1).}
\label{fig:f1}
\end{figure}

\begin{figure}
\epsscale{0.8}
\plotone{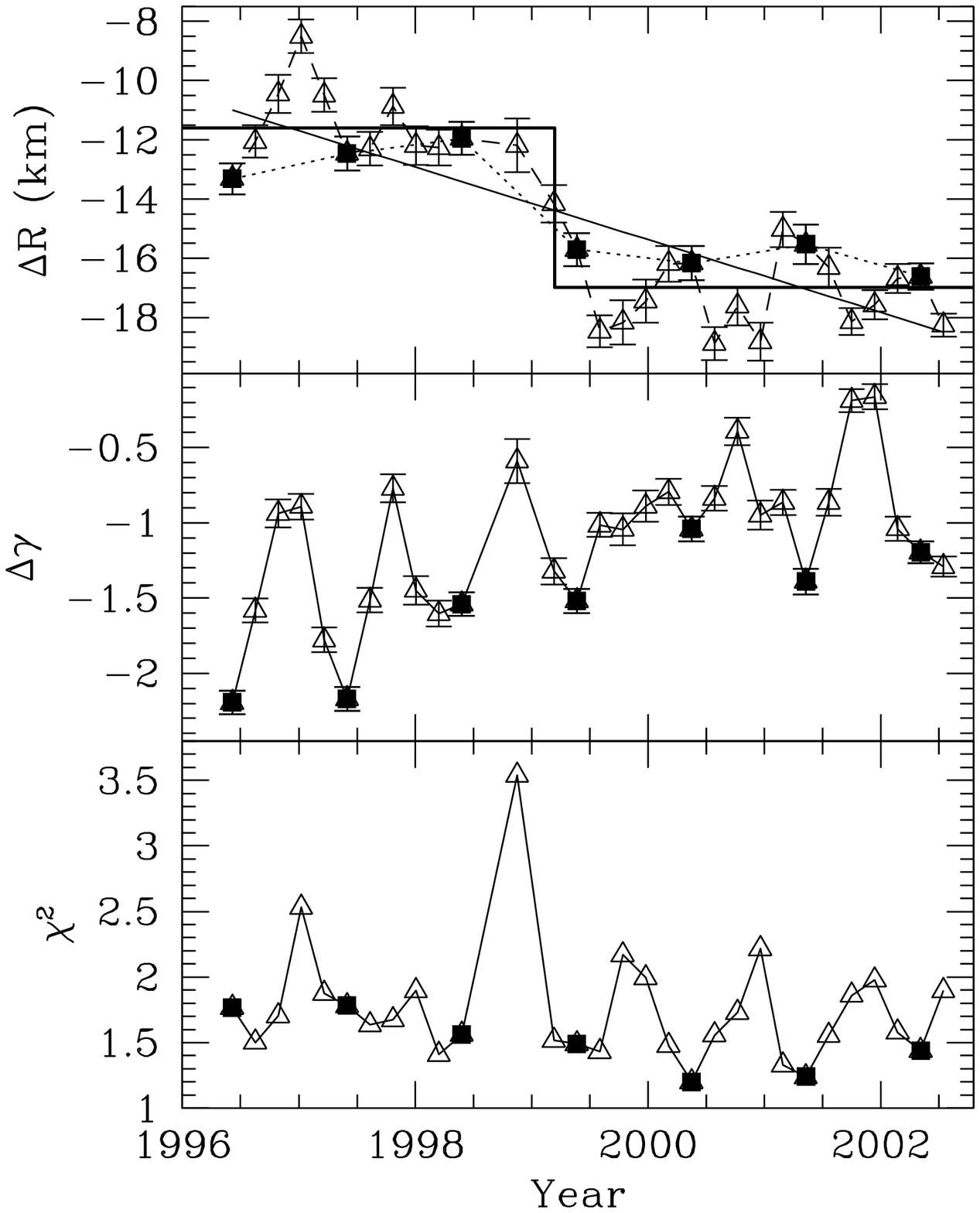}
\caption{The estimated variation in the solar radius,
$\Delta R$,
and the surface term, $\Delta\gamma$ from f-mode frequencies, obtained by fitting
Eq.~1 to frequency difference between a given MDI set and
a solar model. The $\chi^2$ per degree of freedom for each set is
shown in the lowest panel. In each panel the filled squares are the results for
data sets at an interval of 360 days for which the fit is
relatively good. The solid line in the top panel is a straight line
fit to all points, similar to that obtained by DGS. The dashed line
connects all points in upper panel, while the dotted line connects
the filled squares. The heavy line shows a step function fit to
all points with discontinuity at 1999.2.
}
\label{fig:f2}
\end{figure}

\begin{figure}
\plottwo{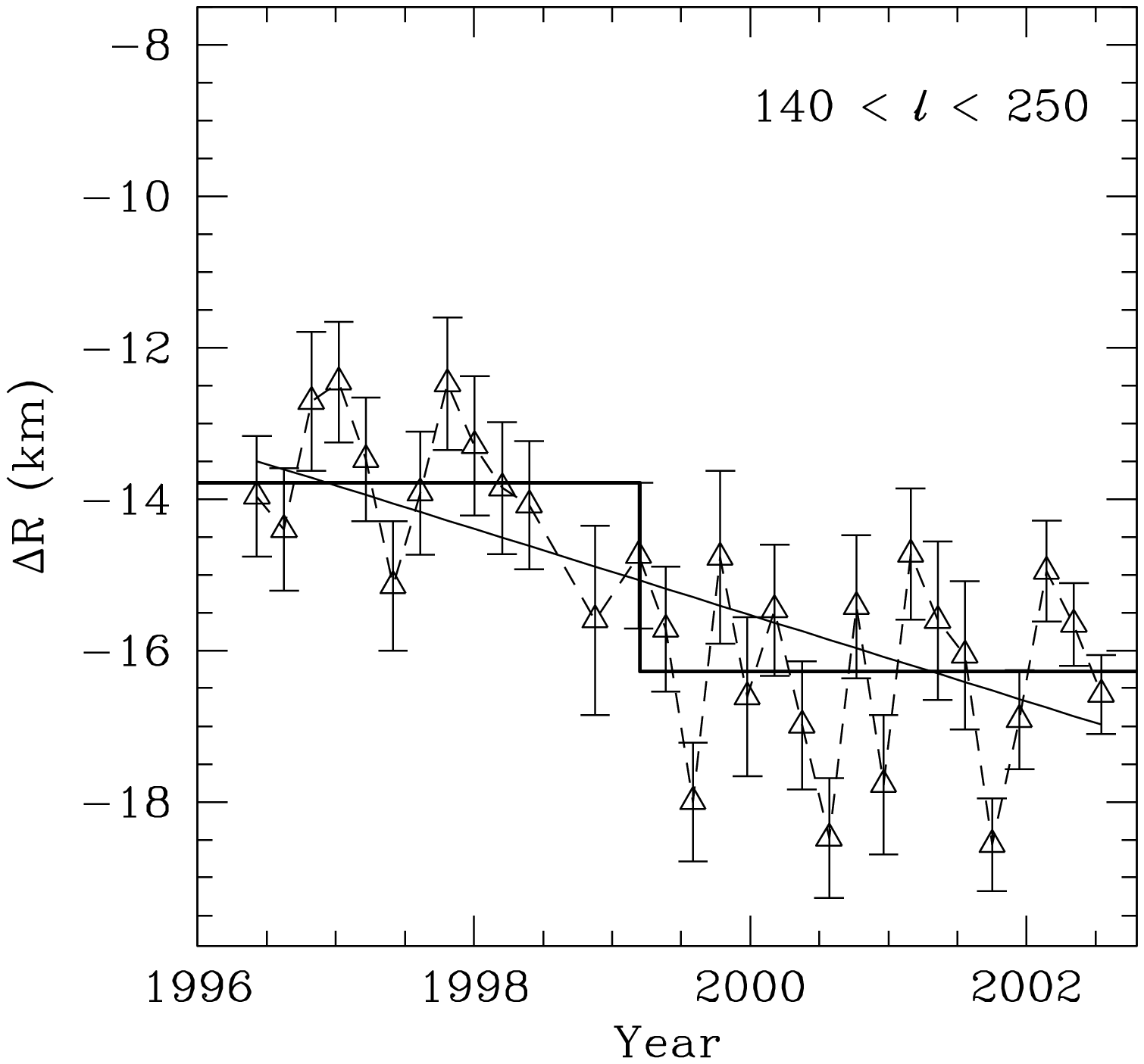}{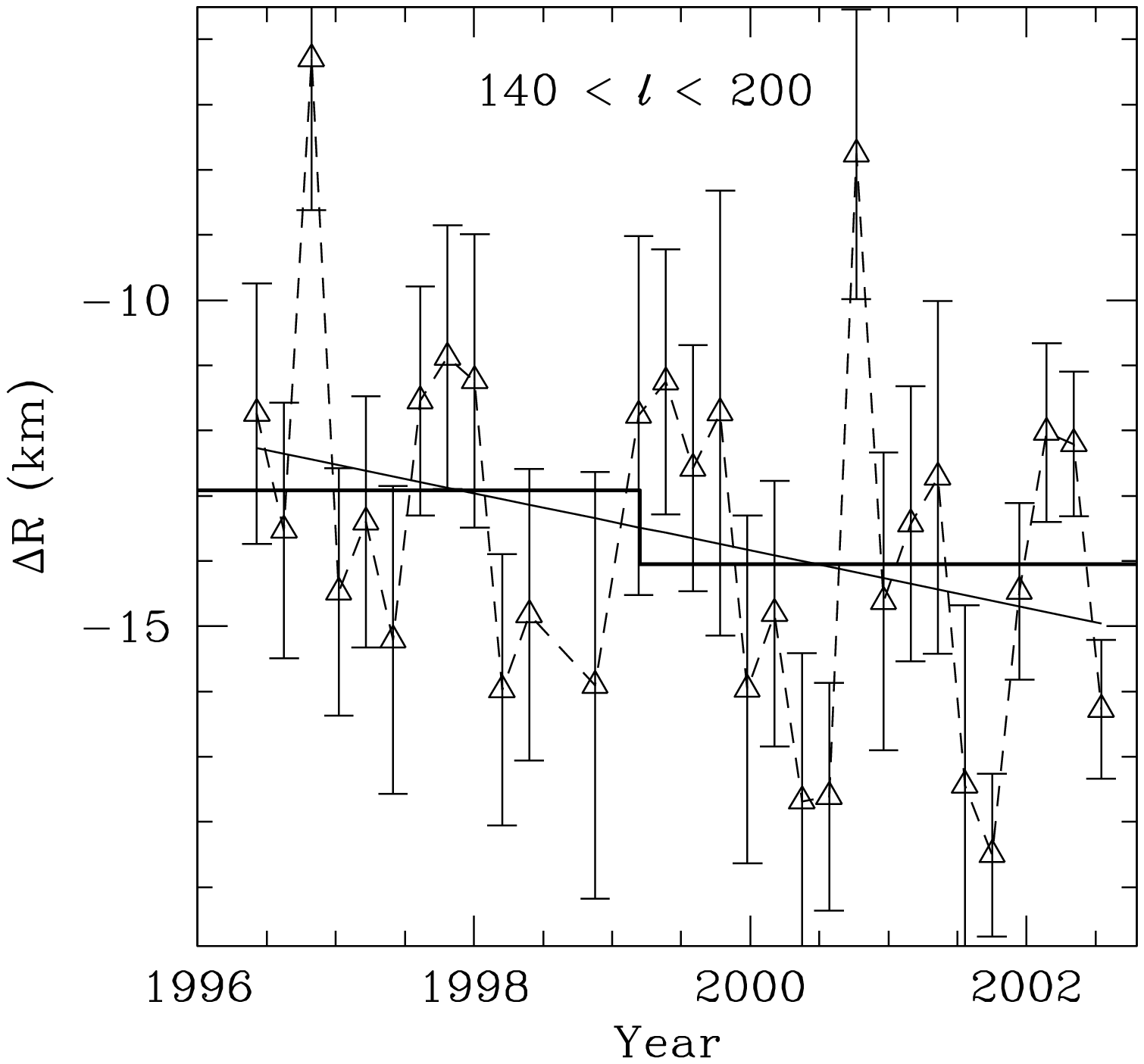}
\caption{The estimated variation in the solar radius,
$\Delta R$, from f-mode frequencies, obtained by fitting
Eq.~1 to frequency difference between a given MDI set and
a solar model. The left and right panels respectively, show the
results when modes with $\ell<250$ and $\ell<200$ are used.
In both panels the solid line shows the linear fit to all points
and heavy line shows a fit to step function.}
\label{fig:f3}
\end{figure}

\begin{figure}
\plotone{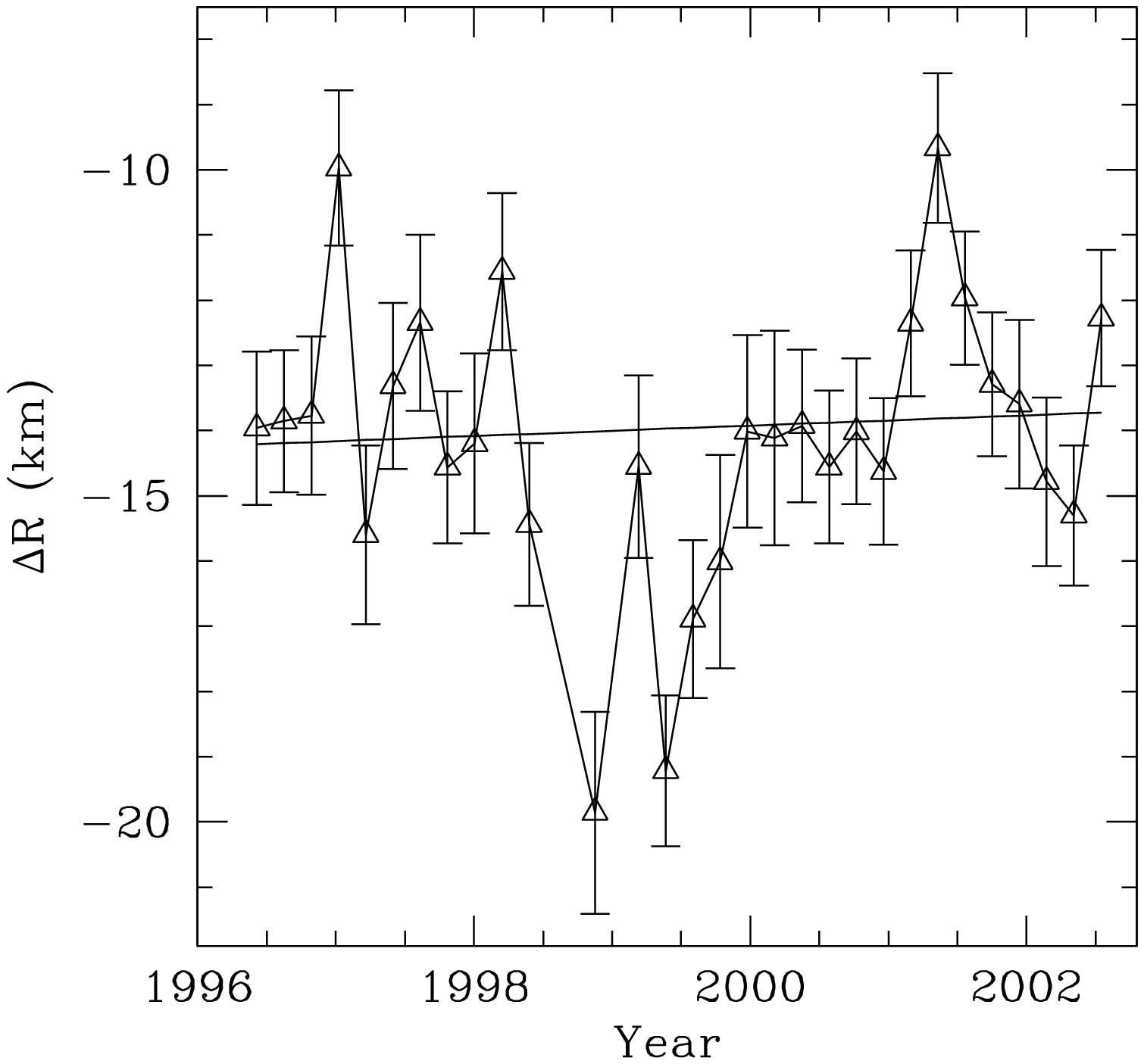}
\caption{The estimated variation in the solar radius,
$\Delta R$, from f-mode frequencies, obtained by fitting
only the first term in
Eq.~1 to frequency difference between a given MDI set and
a solar model. Only modes with $\ell<120$ from the revised
MDI data sets are used. 
The solid line shows the linear fit to all points.}
\label{fig:f4}
\end{figure}

\begin{table}[ht]
\caption{Radius variation as inferred from MDI data}
\bigskip
\begin{tabular}{cccc}
\tableline
Time & 10.7 cm radio flux& $\Delta R$ & $\chi^2$\\
&(sfu)&(km)\\
\tableline
1996.4&\phantom{0}72.4&$-13.3\pm0.5$&1.76\\
1997.4&\phantom{0}74.7&$-12.5\pm0.6$&1.78\\
1998.4&108.5&$-12.0\pm0.6$&1.56\\
1999.4&148.0&$-15.7\pm0.6$&1.49\\
2000.4&186.3&$-16.2\pm0.6$&1.20\\
2001.4&162.4&$-15.5\pm0.6$&1.24\\
2002.4&184.3&$-16.6\pm0.5$&1.44\\
\tableline
\end{tabular}
\end{table}

\begin{table}[ht]
\caption{The rate of radius variation as inferred from MDI data
restricted to different range of degree, $\ell$ and time interval}
\bigskip
\begin{tabular}{ccccc}
\tableline
Time interval& \multicolumn{4}{c}{Inferred rate of radius variation
 (km yr$^{-1}$)}\\
&$\ell<160$&$140<\ell<200$&$140<\ell<250$&$140<\ell$\\
\tableline
\multicolumn{5}{c}{Using both terms in Eq.~(1)}\\
1996.4--2002.6&\nodata&$-0.47\pm0.17$&$-0.57\pm0.08$&$-1.22\pm0.06$\\
1996.4--1998.6&\nodata&$-1.16\pm1.01$&$-0.01\pm0.41$&$-0.16\pm0.28$\\
1999.2--2002.6&\nodata&$-0.52\pm0.40$&$-0.00\pm0.17$&$-0.30\pm0.12$\\
%1997.4--2000.4&\nodata&$-0.81\pm0.60$&$-1.60\pm0.25$&$-2.14\pm0.17$\\
1996.4--2002.6 (revised)&$+0.11\pm0.16$&$-0.50\pm0.11$&$-0.64\pm0.06$&$-1.13\pm0.04$\\
\noalign{\medskip}
\multicolumn{5}{c}{Using only the first term in Eq.~(1)}\\
1996.4--2002.6&$-0.74\pm0.09$&$-1.35\pm0.05$&$-1.94\pm0.04$&$-2.17\pm0.03$\\
1996.4--1998.6&$-0.46\pm0.61$&$-0.50\pm0.25$&$-0.57\pm0.17$&$-0.62\pm0.15$\\
1999.2--2002.6&$-0.03\pm0.22$&$-0.54\pm0.11$&$-0.91\pm0.08$&$-1.11\pm0.07$\\
%1997.4--2000.4&$-1.51\pm0.37$&$-2.26\pm0.15$&$-3.21\pm0.10$&$-3.65\pm0.09$\\
1996.4--2002.6 (revised)&$-0.70\pm0.05$&$-1.29\pm0.04$&$-1.68\pm0.03$&$-1.83\pm0.03$\\
\tableline
\end{tabular}
\end{table}

\clearpage

\begin{thebibliography}{}

\bibitem[\protect\astroncite{Antia}{1998}]{ant98}
Antia, H.  M. 1998, A\&A, 330, 336

\bibitem[\protect\astroncite{Antia}{2002}]{a02}
Antia, H.  M. 2002, in Proc. IAU Coll.\ 188: Magnetic Coupling of
the Solar Atmosphere, ESA SP-505, ed.\ H. Sawaya-Lacoste, p.~71 (astro-ph/0208339)

\bibitem[\protect\astroncite{Antia}{2000}]{ant00}
Antia, H. M., Basu, S., Pintar, J., \& Pohl, B. 2000, Solar Phys., 192, 459

\bibitem[\protect\astroncite{Antia}{2001}]{ant01}
Antia, H. M., Basu, S., Pintar, J., \& Schou, J. 2001, 
in Proc. SOHO 10/GONG 2000 Workshop on Helio- and Astero-seismology
at the Dawn of the Millennium, ESA SP-464, Ed. A. Wilson, p.~27

\bibitem[\protect\astroncite{Antia}{2003}]{ant02}
Antia, H. M., Chitre, S. M., \& Thompson, M. J. 2003, A\&A, 399, 329

\bibitem[Balmforth et al.(1996)]{bal96}
Balmforth, N. J., Gough, D. O., \& Merryfield, W. J. 1996, MNRAS, 278, 437

\bibitem[Basu \& Antia(2000)]{bas00}
Basu, S., \& Antia, H. M. 2000, Solar Phys., 192, 449

\bibitem[Basu \& Antia(2000)]{bas02}
Basu, S., \& Antia, H. M. 2002, in  Proc. SOHO12/GONG+ 2002,
Local and Global Helioseismology: The Present and Future,
ESA SP-517 (Noordwijk: ESA)

%\bibitem[Brown \& JCD(1998)]{bro98}
%Brown, T. M., \& Christensen-Dalsgaard, J. 1998, ApJ, 500, L195

\bibitem[Campbell \& Roberts(1989)]{cam89}
Campbell, W. R., \& Roberts, B. 1989, ApJ, 338, 533

\bibitem[Chitre et al. (1998)]{chi98}
Chitre, S. M., Christensen-Dalsgaard, J., Thompson, M. J. 1998,
in Structure and Dynamics of the Sun and Sun-like Stars,
eds.\ S. G. Korzennik \& A. Wilson, ESA SP-418
(Noordwijk: ESA), p 141

\bibitem[Dappen(1983)]{dap83}
D\"appen, W. 1983, A\&A, 124, 11

\bibitem[Delache et al.(1985)]{del85}
Delache, P., Laclare, F., \& Sadsaoud, H. 1985, Nature, 317, 416

\bibitem[\protect\astroncite{dzi}{1998}]{dzi98}
Dziembowski, W. A., Goode, P. R., DiMauro, M. P.,
Kosovichev, A. G., \& Schou, J. 1998, ApJ, 509, 456

\bibitem[\protect\astroncite{Dziembowski et al}{2000}]{DG200}
Dziembowski, W. A., Goode, P. R., Kosovichev, A. G., \& Schou, J. 2000,
ApJ, 537, 1026

\bibitem[\protect\astroncite{Dziembowski et al.}{2001}]{dgh01}
Dziembowski, W. A., Goode, P. R., \& Schou, J. 2001, ApJ, 553, 897 (DGS)

\bibitem[Emilio et al.(2001)]{MDI2001}
Emilio, M., Kuhn, J. R., Bush, R. I., \& Scherrer, P. 2000, ApJ, 543, 1007

\bibitem[fiala et al.(1994)]{fia94}
Fiala, A. D., Dunham, D. W., \& Sofia, S. 1994,  Solar Phys., 152, 97

\bibitem[Gough(1981)]{dog81}
Gough, D. O. 1981, in Sofia S., ed., Variations of the Solar Constant,
NASA CP 2191, p. 185, Washington D.C.

\bibitem[Gough(2001)]{dog01}
Gough, D. O. 2001, Nature, 410, 313

\bibitem[Laclare et al.(1996)]{lac96}
Laclare, F., Delmas, C., Coin, J. P., \& Irbah, A. 1996,
Solar Phys.,  166, 211

\bibitem[Mecherikunnel(1994)]{mec94}
Mecherikunnel, A. T. 1994, Solar Phys., 155, 211

\bibitem[noeel(1997)]{noe97}
No\"el, F., 1997,  A\&A, 325, 825


\bibitem[\protect\astroncite{schou}{1999}]{sch99}
Schou, J. 1999, ApJ, 523, L181

\bibitem[\protect\astroncite{schou}{1997}]{sch97}
Schou, J., Kosovichev, A.  G., Goode, P.  R., \&
Dziembowski, W. A., 1997, ApJ, 489, L197

\bibitem[\protect\astroncite{trip}{1999}]{tri99}
Tripathy, S. C., \& Antia, H. M., 1999, Solar Phys., 186, 1

\bibitem[Wittmann et al.(1993)]{wit93}
Wittmann, A. D., Alge, E., \& Bianda, M. 1993, Solar Phys., 145, 205



\end{thebibliography}
\end{document}